\documentclass[twocolumn,showpacs,preprintnumbers,amsmath,amssymb]{revtex4}

\usepackage{graphicx}
\usepackage{dcolumn}
\usepackage{bm}

\begin{document}


\title{Collision of One-Dimensional Nonlinear Chains}

\author{Shin-ichiro Nagahiro}
\author{Yoshinori Hayakawa}%
\affiliation{%
Department of Physics, Tohoku University, Sendai, Japan
}%

\date{\today}

\begin{abstract}
We investigate one-dimensional collisions of unharmonic 
chains and a rigid wall. We find that the coefficient of restitution (COR) 
is strongly dependent on the velocity of colliding chains and has a minimum 
value at a certain velocity. The relationship between COR and collision velocity 
is derived for low-velocity collisions using perturbation 
methods. We found that the velocity dependence is characterized by 
the exponent of the lowest unharmonic term of interparticle potential energy.
\end{abstract}

\pacs{45.50.Tn, 05.45.-a, 46.40.-f}
\maketitle

\section{Introduction}
In collisions between two bodies which have internal degrees of freedom,
some of the initial translational energy is transformed into internal energy of
the two bodies. This is the major cause of energy dissipation. To characterize
the macroscopic features, a phenomenological parameter, the coefficient of restitution (COR), 
\begin{eqnarray}
 \eta =\frac{K_r}{K_i}\label{eq:defOfCOR},
\end{eqnarray} 
where $K_i$ and $K_r$ are translational kinetic energy before and after
the collision, respectively, is commonly used. Recent studies 
of collisions are mainly focused on the determination of $\eta$ from 
microscopic mechanisms. 

Hertz developed the theory of collision between frictionless elastic 
bodies \cite{Landau,Timo} based on his static theory of elastic contact \cite{Hertz}. 
In the theory, it is assumed that in low-velocity collisions, the deformation of 
colliding bodies is given by the static theory and the 
production of vibration is totally ignored. Hence, the theory gives 
no information on energy dissipation.  
Plastic deformation is one of the possible means of kinetic
energy dissipation during collisions. Taking this into account, the 
energy dissipation rate $1-\eta$ is found to increase with collision 
velocity with a power law with the exponent $1/2$ \cite{Johnson}. 
Considering viscoelastic properties, the dissipation rate increases 
with the exponent $2/5$ \cite{Kuwabara,Thomas,Roza}.
These results were confirmed experimentally \cite{Raman,Sondergaard,Labous}.  

The results presented above  are based on quasi-static approximation, 
hence it is expected that they are restricted to low-velocity collisions.   
We believe that more general results will be
obtained through microscopic simulations \cite{Gotz,Gerl,HH}. 

Sugiyama and Sasaki \cite{Sugiyama} and Basile and Dumont \cite{Basile} 
considered collisions between simple one-dimensional chains and a rigid wall. 
A chain is composed of $n$ identical 
point particles which interact with nearest-neighbor
particles. If we choose linear force as the interaction between the
particles, the COR is independent of collision velocity and approaches 
unity in the  thermodynamic limit. 

For collision between metallic bodies, plastic 
deformation usually occurs even if collision velocity is low. 
The deformation exceeds the elastic regime in which 
Hooke's law remains valid 
\cite{Johnson,Zukas}. 
Hence nonlinear effects are important in realistic collisions.
In this study, we choose nonlinear force as the interparticle 
interaction and perform the one-dimensional simulations of collision 
between a ``nonlinear chain'' and a rigid wall. We find that the COR of this 
collision has a minimum value at a certain velocity and derive the velocity dependence 
of the COR for low-velocity collisions using perturbation methods.

This paper is organized as follows. In the next section, we briefly
review the studies of Sugiyama and Sasaki \cite{Sugiyama} and Basile and Dumont \cite{Basile}. 
In section 3, we present our results for collision between 
nonlinear chains and a wall. Finally, we summarize our results.  

\section{Collisions of one-dimensional harmonic chains}
\begin{figure}[b]
  \includegraphics[height=5cm,keepaspectratio]{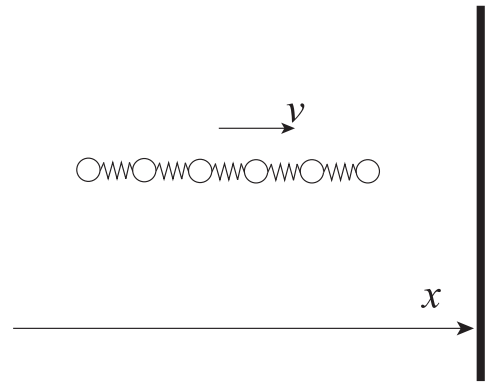}
\caption{\label{fig:model} Schematic of a one-dimensional chain and a rigid wall.}
\end{figure}

First we briefly discuss the collision of one-dimensional harmonic  
chains (model A) with a rigid wall as discussed by
Sugiyama and Sasaki~\cite{Sugiyama}. Consider a chain composed of
$n$ identical point particles labeled $j=1,2,\cdots n$. 
Each particle in the chain is linked to nearest-neighbor particles with
a Hooke's spring, as illustrated in Fig.\ref{fig:model}. 
The Hamiltonian of this system is
written as
\begin{eqnarray}
 {\cal H}=\frac{m}{2}\sum_{j=1}^n \dot{x}_j^2 + \frac{1}{2}m\omega^2
\sum_{j=1}^{n-1}(x_{i+1}-x_{i}-\ell)^2+V_{w},\label{eq:hamiltonian}
\end{eqnarray}  
where $x_j$ and $\dot{x}_j$ are the position and velocity of the $j$-th
particle, respectively, and $\ell$ is the natural length of the springs. 
We assume the chain is homogeneous and all springs have the same spring constant 
$k(=m\omega^2)$. $V_{w}$ represents the 
hard-core potential of a rigid wall located at $x=0$ where collision takes place. 
During the collision process, only the particle $j=1$, 
which is at the end of the chain, interacts with the wall since the positional 
order of the particles is always kept as $x_n < x_{n-1} < \cdots < x_2 < x_1$. 
In the following discussion, we assume that the wall is so rigid 
that the particle $x_1$ simply reverses its velocity as
$-\dot{x}_1\to \dot{x}_1$.

Although there is no dissipation term in the Hamiltonian, vibration energy 
which remains after the collision is regarded as the ``dissipated'' portion of energy. 
The COR $\eta$ is therefore evaluated as
\begin{eqnarray}
 \eta=1-\frac{E_{vib.}}{E},\label{eq:cor}
\end{eqnarray}  
where $E$ is the total energy and $E_{vib.}$ is the vibration energy 
which remains after the collision.

In the absence of $V_w$, the equation of motion is written as
\begin{eqnarray}
 \ddot{\bf x}=-\omega^2({A}{\bf x}+{\bf b}),\label{eq:eqOfMo}
\end{eqnarray} 
where ${\bf x}=(x_1, x_2, \cdots, x_n)$, ${\bf b}=(-\ell, 0,\cdots, 0, \ell)$
and ${A}$ is an $n\times n$ matrix of the form
\begin{eqnarray}
A=\left(
   \begin{array}{ccccccc}	
     1 & -1 &  0 &  &   & \cdots & 0 \\
    -1 & 2  & -1 &  0  &  & \cdots & 0 \\
     0 & -1 &  2 & -1  &    0   & \cdots & 0 \\
     \vdots & & & \ddots & & & \vdots \\
     0 & & \cdots & & 0 & -1& 1 \\
   \end{array}
   \right).
\end{eqnarray}
Taking a principal-axis coordinate system in which $A$ is diagonal, 
internal vibration of the chain can be represented using $n$ noninteracting 
fundamental modes. 
In the presence of $V_w$, collision between the chain and the wall is realized 
as follows. 
Assume that the particle  $j=1$ collides with 
the wall $f(n)$ times at $t_1, t_2, \cdots, t_{f(n)}$. 
The fundamental modes describe equi-energy elliptical orbits in phase space. 
The orbits are discontinuous at $t=t_1, t_2, \cdots, t_{f(n)}$.

\begin{figure}[tb]
  \includegraphics[height=6cm,keepaspectratio]{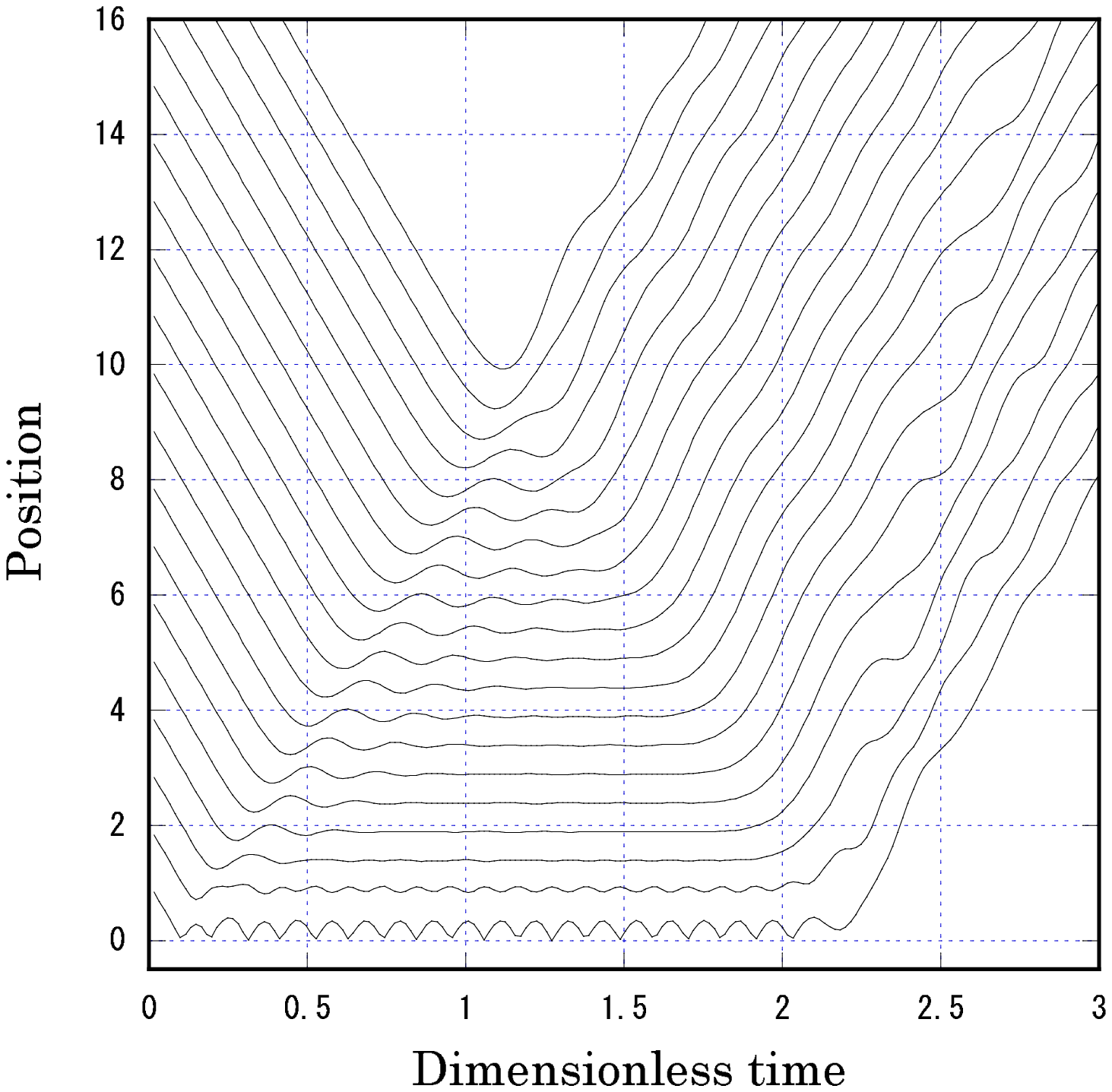}
\caption{\label{fig:collisionWithWall} Trajectories of the particles in the $n=20$ colliding chain. 
(Reduced units where $c_l=\tau=1$.)}
\end{figure}

For the following numerical simulations, we choose initial conditions 
\begin{eqnarray}
 \left\{
  \begin{array}{cccc}
   x_j&=&\ell~j,& \\
   \dot{x}_j&=&-v_0&(j=1,2,\cdots,n).
  \end{array}
 \right.\label{eq:initialState}
\end{eqnarray}
Before collision, the chain has no internal
vibration, {\it i.e.,} zero temperature.
Figure \ref{fig:collisionWithWall} 
shows the collision between the wall and 
the chain with $n=19$.
Each line is a trace of the trajectory of a particle. 
In the plot, units on the time axis are taken as $\tau=(n-1)\ell/c_l$. $\tau$ indicates 
the duration in which the longitudinal sound wave propagates from one end of the chain to the other.
Let us call the time $t_c$ during which the collision takes place 
"contact time". Here we note that contact time is almost 
equal to $2\tau$ for large $n$. COR and contact time are independent of 
the collision velocity of the chain 
because time intervals $\triangle t_q=t_{q+1}-t_{q}$ are independent of 
initial velocity for every $q$ and the orbits retain similar forms 
even if initial velocity of the chain varies. Therefore, $E_j/E$, the ratio of 
the $j$-th fundamental mode energy to total energy, does not 
depend on initial velocity.
The COR, which is a ratio of energy, also does not depend on 
initial velocity.
Hereafter, we use the reduced unit $c_l=\tau=1$, setting $\ell$, $m$ 
and $k$ as unity.

In model A, the coefficient of restitution $\eta$ depends only on $n$. 
An approximate expression of $\eta$ was derived by
Basile and Dumont\cite{Basile}. 
To simplify the problem, we assume that $\triangle t_q$ and the velocity before the $q$-th collision $v_{q}$ are
constant for every $q$. Numerical simulations show that these are good approximations.
Under this assumption, the vibration energy of the 
$j$-th mode after the collision process is given as
\begin{eqnarray}
 E_j=\frac{4m}{n}v^2\cos^2\left(\frac{\pi j}{2n}\right) \frac{ \sin^2\{ \omega_j f(n) \triangle t/2 \}}
  {\sin^2 \{\omega_j\triangle t/2\}},
\label{eq:energy}
\end{eqnarray}  
where $\omega_j$ is the frequency of the $j$-th mode, which is $\omega_j=\sin(\pi j/n)$. 
$\triangle t$, $f(n)$ and $v$ are determined from numerical
simulations \cite{Basile}. However we can also estimate these values by solving 
the collision of the chain with $n=2$: 
\begin{eqnarray}
 \triangle t &=& \frac{\pi}{\sqrt{2}\omega}=2.22/\omega~~~(2.31/\omega),\label{eq:deltat}\\
 f(n) &\simeq& \frac{2\sqrt{2}}{\pi}n=0.90n~~~(0.867n),\label{eq:fn}\\
 v &=& \frac{\pi}{2\sqrt{2}}v_0=1.11v_0~~~(1.15v_0),\label{eq:v}
\end{eqnarray}  
where the values in brackets were obtained from a numerical simulation 
of an $n=500$ chain. We can estimate an approximate 
value of $\eta$ by using Eqs. (\ref{eq:deltat}), (\ref{eq:fn}) and (\ref{eq:v}).

\begin{figure}[tb]
  \includegraphics[height=6cm,keepaspectratio]{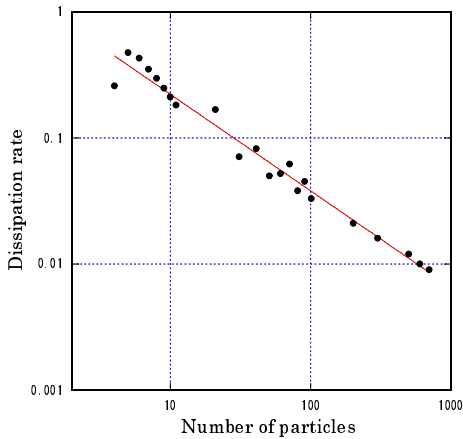}
\caption{\label{fig:COR-N} A log-log plot of dissipation rate versus number of particles. 
The solid line has slope $-2/3$. (Reduced units where $c_l=\tau=1$.) }
\end{figure}

To obtain the asymptotic behavior of $1-\eta$ for
large $n$, we expand the dispersion relation. 
\begin{eqnarray}
 \omega_j= \sqrt{k/m}\frac{\pi j}{n}\left(1-\frac{\pi^2j^2}{24n^2}+\cdots\right)\label{eq:expandedDR}
\end{eqnarray}
Substituting only the leading order of Eq.(\ref{eq:expandedDR}) into Eq.(\ref{eq:energy}), 
we have $\eta=1$. Taking the second order into account, we obtain 
\begin{eqnarray}
 1-\eta \sim 0.652~n^{-\frac{2}{3}}, 
\end{eqnarray}
in the limit $n\to\infty$.
This relationship agrees very well with the numerical
result. In Fig. \ref{fig:COR-N}, we plot the dissipation rate $1-\eta$ 
versus $n$.

\section{Unharmonic chains}
In the case of harmonic chains, COR and contact time were independent of 
initial velocity $v_0$. 
Here, we introduce nonlinearity to the springs in the chain (model B). 

In order to maintain a universal viewpoint, we first consider a
velocity scale characterized by the nonlinearity of
the spring. Let $U(x)$ be the potential energy of the spring, which 
is chosen to be $U(1)=0$ assuming the natural length of the spring to be unity. 
We can define the amplitude $x^*$ at which the 
harmonic term and the sum of the remaining terms are equal in the Taylor expansion, 
{\it i.e.,} $x^*$ is given by the solution of 
\begin{eqnarray*}
 \frac{1}{2}\left.\frac{d^2 U(x)}{dx^2}\right|_{x=1}({x^{*}-1})^2=U(x)-\frac{1}{2}\left.\frac{d^2 U(x)}{dx^2}\right|_{x=1}({x^{*}-1})^2.
\end{eqnarray*}   
The velocity which corresponds to  $x^*$ is 
\begin{eqnarray*}
 {v^*}=\sqrt{2U(x^*)}.
\end{eqnarray*}
Hereafter, we discuss the velocity dependence of the collisions on the scale of $v^*$.

Let us choose the following three types of potential and compare the results of simulation.
 
(a) Lennard-Jones potential
\begin{eqnarray*}
  U_{LJ}(x)=\frac{1}{72}\left\{\left(\frac{1}{x}\right)^{12}-2\left(\frac{1}{x}\right)^6\right\} \label{eq:LJ}
\end{eqnarray*} 

(b) Toda potential
\begin{eqnarray*}
  U_{toda}(x)&=&\frac{a}{b}e^{-b(x-1)}+a(x-1) \label{eq:toda}
\end{eqnarray*}

\begin{figure}[tb]
  \includegraphics[height=6cm,keepaspectratio]{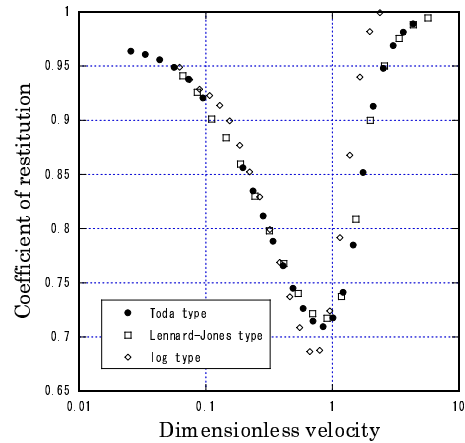}
  \caption{\label{fig:cor-v} Coefficient of restitution for the collisions of log-, Toda- and 
Lennard-Jones-type chains as a function of collision velocity. The chains consist of 100 particles.} 
\end{figure}

(c) Log-type potential
\begin{eqnarray*}
  U_{log}(x)&=&{x}-\log\left({x}\right) \label{eq:log}
\end{eqnarray*}
For the Toda potential, we set $ab=1$ and $b=10$. 
Each potential has one minimum at $x=1$ and the function forms are similar. 
However, increasing behaviors of the repulsive forces derived from the three potentials 
are, in a very short distance, different from one another.  
In Fig. \ref{fig:cor-v}, we plot the COR versus initial velocity for each case,
as determined by numerical simulations. 
In the limit of small velocity, it is clear that COR approaches 
the value obtained for the harmonic chain. The COR decreases with 
increasing initial velocity and has a minimum near $v=1$. 
We note that the COR lies on almost the same curve for all three types of potential. 

\begin{figure}[bt]
  \includegraphics[height=6cm,keepaspectratio]{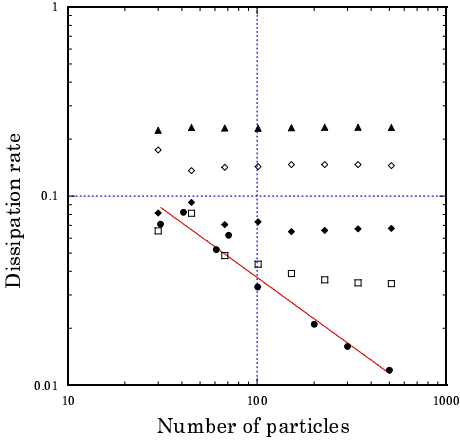}
  \caption{\label{fig:cor-Nv} The relationship between dissipation rate and number of particles in the collision of 
Lennard-Jones-type chain. Each plot corresponds to a different initial velocity: $v_0=0.780~(
\blacktriangle)$, $0.390~(\lozenge)$, $0.160~(\blacklozenge)$, $0.078~(\square)$ and $0.008~(\bullet)$.}
\end{figure}

Figure \ref{fig:cor-Nv} shows $n$-dependence of $\eta$ for different values of initial 
velocity. In very low-velocity collision, as expected from the result for model A, dissipation 
rate $1-\eta$ will approach to zero in the thermodynamic limit. 
Near ${v}_0=1$, the COR does not approach unity but remains a 
constant less than unity even in the thermodynamic limit. 
We can hence conclude that nonlinearity 
of the potential $U(x)$ causes the dissipation, which does not appear in model A.

Using a technique based on the perturbative theory,
we consider the collision of model B for small initial incident 
velocity $v_0\ll 1$.
During collision processes, the particle $j=1$ 
transmits vibration force to its neighbor 
particle. 
Let us regard this force as external force $F(t)$ which 
acts on the chain. The characteristic frequency of this force is 
 $\omega_{ext}=2\pi/\triangle t \simeq 2\sqrt{2}\omega$. This 
frequency is higher than any frequency of 
fundamental modes of the chain. Hence, no fundamental mode 
is excited by the force. In this situation, the amplitude 
of each particle's vibration is rapidly damped progressively 
into the chain, $i.e.,$ the particle 
$j=1$ has the largest amplitude during collision.
It is expected that $\triangle t$ will shorten with increasing 
initial velocity.
As a first approximation, we take into account only 
the change of $\triangle t$ against initial 
velocity $v_0$ in Eq. (\ref{eq:energy}).
To estimate the velocity dependence of $\triangle t$, let us consider the
collision of a Lennard-Jones-type chain with $n=2$. In this case, particle $j=1$
collides with the wall two times. The Hamiltonian is
\begin{eqnarray}
 {\cal H}=\frac{m}{2}(\dot{x}^2_1+\dot{x}^2_2)+U_{LJ}(x_2-x_1-1)+V_{w}.
\end{eqnarray} 
Let $x=x_2-x_1-\ell$ and $x_g=(x_2+x_1)/2$. For initial conditions, we adopt 
Eq. (\ref{eq:initialState}). Immediately after the first collision of
particle $x_1$, $x$ and $x_g$ and their time derivatives become
\begin{eqnarray}
 \left\{
  \begin{array}{cccc}
   x       &=&   0   & \\
   \dot{x} &=& -2v_0 &,~~~~~~~
  \end{array}
 \right. 
 \left\{
 \begin{array}{cccc}
    x_g    &=& \ell/2 & \\
   \dot{x}_g &=& 0      & .
 \end{array}
 \right.\label{eq:initialN2}
\end{eqnarray}
Taking into account the second order of $U_{LJ}$ and solving the
equation of motion with the initial conditions of Eq. (\ref{eq:initialN2}) in
the first-order perturbation theory, one can obtain the interval $\triangle t$
between first and second collisions of particle $x_1$ as
\begin{eqnarray}
 \triangle t \simeq \frac{\pi}{\sqrt{2}}\left(1-\alpha \frac{v_0}{v^*}\right),\label{eq:contactTime}
\end{eqnarray}
with $\alpha=0.808$. Substituting this into Eq. (\ref{eq:energy}), we have 
\begin{eqnarray}
 E_j=\frac{4}{n}\cos^2\left(\frac{\pi j}{2n}\right)
  \frac{\sin^2[\frac{1}{2}\omega_jf(n)\triangle t(1-\alpha \tilde{v})]}
  {\sin^2[\frac{1}{2}\omega_j\triangle t(1-\alpha \tilde{v})]}.
\label{eq:energyV}
\end{eqnarray} 
In the limit $n\to \infty$, $\omega_j$ 
can be replaced with the linear form 
\begin{eqnarray}
 \omega_j=\omega\frac{\pi j}{n}.
\end{eqnarray}
When $v/c_l\ll 1$, Eq. (\ref{eq:energyV}) can be approximated as
\begin{eqnarray*}
 E_j\simeq \frac{32}{n\pi^2}\left(\frac{n}{j}\right)^2\sin^2 \left[ \alpha v \pi n \left(\frac{j}{n}\right) \right].
\end{eqnarray*}
Substituting this into Eq. (\ref{eq:cor}), we have the dissipation rate 
\begin{eqnarray}
 1-\eta=\sum_{j=1}^{n}\frac{E_j}{E}&\simeq&\frac{64}{n\pi^4}\int_{0}^{1} dx \frac{1}{x^2}\sin^2(\alpha vnx) \nonumber\\ 
 &\simeq& \frac{64}{\pi^3}\alpha v \int_{0}^{\infty} dx \frac{\sin^2(\alpha x)}{x^2}\nonumber\\
 &=& Cv, \label{eq:COR-v}
\end{eqnarray}
where $C=\frac{32}{\pi^2}\alpha \simeq 2.61$.
This implies that the dissipation rate $1-\eta$ increases with 
the power law $v^p$ where $p=1$. This result agrees with numerical simulation 
for low-velocity collision. However the constant $C$ is not in 
accord with the above result (our numerical simulation gives $C=0.66$).
\begin{figure}[bt]
  \includegraphics[height=6cm,keepaspectratio]{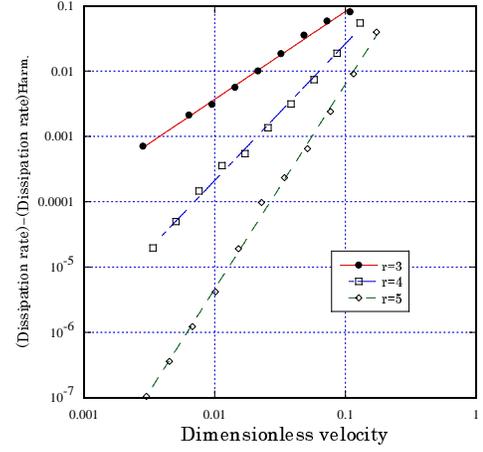}
  \caption{\label{fig:cor-vn} Increase of dissipation rates. 
We plot the dissipation rate of model B minus that of model A for $r=3, 4$ and $5$.
Fitted lines have slopes $1.34, 2.09$ and $3.14$, respectively.}
\end{figure}

The above result directly depends on the exponent of the 
lowest un-harmonic term of interparticle potential energy. In the case of 
Lennard-Jones potential, the exponent $r=3$. The exponent is the same for 
other types of chains. 
Let us discuss more general cases. Suppose that the interparticle potential 
energy can be expanded around its equilibrium position as 
\begin{eqnarray}
 U(x)=\frac{1}{2}x^2+cx^r+{\rm ~higher~order},\label{eq:nth-orderPotential}
\end{eqnarray}
where the constant $c$ is a positive (negative) number when $r$ is
even (odd). In this case,  the contact 
time shortens as $v_0^{r-2}$. 
The dissipation rate hence increases as  
\begin{eqnarray}
 1-\eta \propto \left(\frac{v}{v^*}\right)^{r-2}.\label{eq:dissRate}
\end{eqnarray}
Taking potential energy Eq.(\ref{eq:nth-orderPotential}) for the 
interaction of particles in model B, we plot numerical results of 
COR in Fig. \ref{fig:cor-vn} for $v \ll 1$.  
The results agree with Eq. (\ref{eq:dissRate}). 

When the particles in the chain can be regarded as rigid spheres, interparticle 
interaction is of the $\delta$ function type. 
Collision between the two spheres is reduced to a simple exchange of their momenta. 
We can exactly solve the entire dynamics of the collision between the chain and 
a rigid wall. The sphere $j=1$ collides with the wall $n$ times and 
no internal vibration remains after the collision, {\it i.e.,} the COR is exactly 
unity in this case. 
\begin{figure}[bt]
  \includegraphics[height=6cm,keepaspectratio]{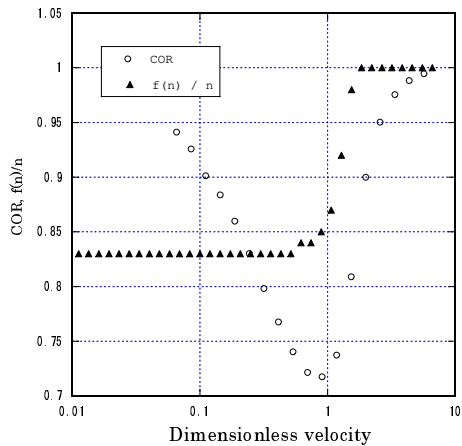}
  \caption{\label{fig:CCT} Plot of ratio of $f(n)$ to $n$ and COR versus velocity 
for Lennard-Jones-type chain. Both increase at $\tilde{v}\approx 1$.}
\end{figure}

In the limit of high-velocity collision of model B, particles interact like 
rigid spheres because $U_{LJ}$, $U_{toda}$ and $U_{log}$ all behave like hard core 
potentials within a very short distance. In Fig. \ref{fig:CCT}, we show the COR of 
a Lennard-Jones-type chain and $f(n)/n$ versus collision velocity on the same plot. 
The rate $f(n)/n$ approaches unity for ${v}>1$. 
This indicates that particles interact like rigid spheres in high-velocity 
collision. Consequently COR increases in the high-velocity regime.
This is a feature which only the one-dimensional model exhibits and is 
unrealistic. In real systems, plastic deformation is crucial 
in such high-velocity collisions. 

\section{Summary}
We have presented a simple one-dimensional microscopic model of colliding bodies to 
understand the energy dissipation process. Lennard-Jones, Toda and Log-type potentials 
are chosen as interactions between particles. We found that 
the COR depends on the initial velocity and is minimum at
$v/v* \simeq 1$. These behaviors are independent of the potential form. 
In low-velocity collisions, the relationship between 
the energy dissipation rate and collision velocity is derived using perturbation methods.

\end{document}